\documentclass{article}

\usepackage[preprint,nonatbib]{neurips_2021}

\usepackage[utf8]{inputenc} 
\usepackage[T1]{fontenc}    
\usepackage{hyperref}       
\usepackage{url}            
\usepackage{booktabs}       
\usepackage{amsfonts}       
\usepackage{nicefrac}       
\usepackage{microtype}      

\usepackage[pdftex]{graphicx}
\usepackage{balance} 
\usepackage{multirow}
\usepackage[ruled,vlined]{algorithm2e}
\SetKwBlock{Repeat}{repeat}{}
\usepackage{color,soul}	
\usepackage[font=footnotesize,labelfont=bf,skip=1pt]{caption}
\usepackage{subcaption}
\usepackage{bbm}
\usepackage{bm}
\usepackage{amsmath,amssymb}
\usepackage{amsmath}
\usepackage{cleveref}
\usepackage{wrapfig}

\usepackage{tikz}

\newcommand{\BibTeX}{\rm B\kern-.05em{\sc i\kern-.025em b}\kern-.08em\TeX}

\DeclareMathOperator{\EX}{\mathbb{E}}

\usepackage[normalem]{ulem}

\title{Disentangling Successor Features for Coordination in Multi-agent Reinforcement Learning}

\author{%
  Seung Hyun Kim \\
  University of Illinois at Urbana-Champaign\\
  Champaign, IL \\
  \texttt{skim449@illinois.edu} \\
  \And
  Neale Van Stralen \\
  University of Illinois at Urbana-Champaign\\
  Champaign, IL \\
  \texttt{nealeav2@illinois.edu} \\
  \And
  Girish Chowdhary \\
  University of Illinois at Urbana-Champaign\\
  Champaign, IL \\
  \texttt{girishc@illinois.edu} \\
  \And
  Huy T. Tran \thanks{corresponding author} \\
  University of Illinois at Urbana-Champaign\\
  Champaign, IL \\
  \texttt{huytran1@illinois.edu} \\
}

\begin{document}

\maketitle

\begin{abstract}

Multi-agent reinforcement learning (MARL) is a promising framework for solving complex tasks with many agents. However, a key challenge in MARL is defining private utility functions that ensure coordination when training decentralized agents. This challenge is especially prevalent in unstructured tasks with sparse rewards and many agents. We show that successor features can help address this challenge by disentangling an individual agent's impact on the global value function from that of all other agents. We use this disentanglement to compactly represent private utilities that support stable training of decentralized agents in unstructured tasks. We implement our approach using a centralized training, decentralized execution architecture and test it in a variety of multi-agent environments. Our results show improved performance and training time relative to existing methods and suggest that disentanglement of successor features offers a promising approach to coordination in MARL.

\end{abstract}


\section{Introduction}

Multi-agent reinforcement learning (MARL) has shown promising results for applications including traffic control \cite{Ma2020}, smart grids \cite{Roesch2020, Ghasemi2020}, autonomous driving \cite{Palanisamy2019}, and UAV control \cite{Yang2018}.
Many existing approaches implement centralized architectures to ensure coordination among agents \cite{Ma2020,Shua2019,Lin2019}.
However, fully centralized approaches struggle when faced with exponentially large state-action spaces and communication constraints. Decentralized execution helps address these issues by allowing agents to make independent decisions without communication.
An open question in MARL is then how to ensure coordination among such decentralized agents; particularly in unstructured tasks with sparse rewards and many agents \cite{Amato2013,matignon2012independent}.

Previous work has addressed this problem by defining private utility functions, often referred to as individual value functions, for optimizing individual agent policies. However, current methods struggle to ensure scalable and stable learning in complex environments like the Starcraft Multi-Agent Challenge (SMAC) \cite{Samvelyan2019}.
Two concepts that underlie these struggles are factoredness and learnability. Factoredness ensures a private utility is aligned with the global utility (e.g., the global value function), while learnability stabilizes training by disentangling an individual agent's impact on the global utility from that of all other agents \cite{Tumer2006}.
Recent frameworks, such as QMIX \cite{Rashid2018}, QTRAN \cite{Son2019}, WQMIX \cite{Rashid2020}, and LICA \cite{Zhou2020}, propose private utilities that achieve high factoredness and show promising results.
However, these methods do not explicitly consider learnability, limiting their performance in highly complex environments with many agents.
The counterfactual multi-agent (COMA) method implements a technique for computing private utilities with high factoredness and learnability \cite{Foerster2018}, but requires a complex central critic to learn a joint action value function, limiting its performance in large multi-agent systems (MAS).

In this work, we ask whether \emph{successor features (SFs) can be used to disentangle an individual agent's impact on the global value function from that of all other agents, and if this disentanglement then enables better training of decentralized agents}.
More specifically, we expect such disentanglement to improve decentralized training by enabling {compact} representation of a private utility that has high factoredness and {high learnability}.
We introduce three key ideas to address this question, all of which leverage the ability of SFs to separate environment dynamics from rewards. First, we introduce a private utility, based on a modified formulation of the estimated difference utility (EDU) \cite{Tumer2006}, that uses SFs to disentangle the impact of an agent from the global value function. We use our private utility, named SF-EDU, as an individual value function for training decentralized agents. Second, we introduce a process for learning the SF disentanglement vector used to calculate our SF-EDU. We optimize this disentanglement for learnability, which we estimate using SFs, to improve training stability and performance. Third, we introduce a centralized training, decentralized execution (CTDE) architecture with a shared feature encoding that efficiently implements these ideas.

We test our approach in a variety of multi-agent environments. Our results suggest that disentanglement of SFs is a promising approach for improving performance in MARL relative to baseline methods, particularly in unstructured tasks with many agents, such as SMAC.

\section{Background}


We model our problem as a decentralized partially observable Markov decision process (dec-POMDP), which is defined by the tuple $(\mathcal{I},\mathcal{S},\mathcal{O}^i,\mathcal{A}^i,\mathcal{P},R)$. Here, $\mathcal{I}$ is a finite set of agents, $\mathcal{S}$ is the state space of the environment, $\mathcal{O}^i$ is the observation space of agent $i \in \mathcal{I}$, and $\mathcal{A}^i$ is the action space of agent $i$.
At every time step, agents execute a joint action $\textbf{a} = (a^1, ..., a^{|\mathcal{I}|})$ where $\textbf{a}$ is in the joint action space $\mathcal{A} = \mathcal{A}^1 \times  ... \times \mathcal{A}^{|\mathcal{I}|}$, the environment updates to a new state $s$ based on the transition function $\mathcal{P}(s' | s,\textbf{a}): \mathcal{S}\times \mathcal{A} \times \mathcal{S} \xrightarrow[]{} [0,1]$, and a global reward $r$ is given to the team based on the reward function $R (s, \textbf{a}): \mathcal{S} \times \mathcal{A} \xrightarrow[]{} \mathbb{R}$.

We assume each agent has access to its own observation history $\tau^i_t=[o^i_1,..., o^i_{t}]$, where $o^i \in \mathcal{O}^i$ and $\tau^i_t \in T^i$.
Our goal is to optimize a stochastic policy for each agent, $\pi^i(a^i|\tau^i_t): T^i \times \mathcal{A}^i \xrightarrow[]{} [0, 1]$, such that the global state value $V^\pi_G(s) = \EX_{\pi}(\sum_{l=0}^{\infty}\gamma^l r_{t+l}| s_t = s)$ or the global action value $Q^\pi_G(s,\textbf{a}) = \EX_{\pi}(\sum_{l=0}^{\infty}\gamma^l r_{t+l} | s_t=s, \textbf{a}_t=\textbf{a})$ of the environment is maximized, where $\gamma \in [0, 1)$ is a discount factor and $\pi$ is a joint policy.

\subsection{Factoredness and Learnability}
The factoredness, $F_{g_i}$, of a private utility $g_i$ is defined as,
\begin{equation}\label{eqn:factoredness_Tumer}
    F_{g_i} = \frac
        {\sum_\textbf{z}\sum_{\textbf{z}'}u \left[(g_i(\textbf{z})-g_i(\textbf{z}'))(G(\textbf{z})-G(\textbf{z}')) \right]}
        {\sum_\textbf{z}\sum_{\textbf{z}'}1},
\end{equation}
where $\textbf{z}$ is a joint move (e.g., a joint state action ($s, \textbf{a}))$, $u[\cdot]$ is the unit step function which has a value of one when the input argument is positive and zero otherwise, and $G$ is the global utility function.
High factoredness implies that positive changes in private utility result in positive changes in global utility, thus ensuring globally beneficial updates in decentralized training.

The learnability, $\lambda_{i,{g_i}}(\textbf{z})$, of $g_i$ is defined as,
\begin{equation}\label{eqn:learnability_Tumer}
    \lambda_{i,{g_i}}(\textbf{z}) = \frac
        {\EX_{z_i'} \left[ |g_i(\textbf{z})-g_i(\textbf{z}_{-i}+z_i')| \right]}
        {\EX_{\textbf{z}_{-i}'} \left[ |g_i(\textbf{z})-g_i(\textbf{z}_{-i}'+z_i)| \right]},
\end{equation}
where $z_i$ are the components of $\textbf{z}$ that only depend on agent $i$ and $\textbf{z}_{-i}$ are the components of $\textbf{z}$ that depend on all agents other than agent $i$. Learnability measures how sensitive an agent's private utility is to its own actions rather than the actions of other agents; high learnability thus reduces noise from other agents when updating an agent's private utility.


The EDU private utility assures full factoredness and promotes learnability by minimizing the influence of other agents on an individual agent's private utility \cite{Tumer2006}.
For a private utility $g_i(\textbf{z})$, the EDU is defined as,
\begin{equation}\label{eqn:EDU_Tumer}
    EDU_i \equiv G(\textbf{z}) - \EX_{z_i}[G(\textbf{z})|\textbf{z}_{-i}],
\end{equation}
where $\EX_{z_i}[G(\textbf{z})|\textbf{z}_{-i}]$ disentangles (or marginalizes) the impact of all possible joint-actions for agents other than agent $i$ from the global utility.
However, implementing EDU is challenging when there is no explicit method for calculating this marginalization term.
Function approximation can help with this challenge \cite{Colby2014,Foerster2018}, but is still limited in environments with many agents or complex interactions among agents.
That is, this term is difficult to approximate if the value function is highly elastic in response to the joint actions of other agents.
We leverage SFs to overcome this challenge.

For the remainder of this paper, we define the global utility $G(\textbf{z})$ as the global state value function $V_G^\pi(s)$ and a private utility $g_i(\textbf{z})$ as an individual state value function $V^{\pi^i}_i (\tau^i_t)$.
Note that we use action value functions instead of state value functions when calculating expectations of utility over actions.
While we refer to $V^{\pi^i}_i$ and $Q^{\pi^i}_i$ as individual value functions for agent $i$, we note that they are more precisely utility functions since they do not strictly estimate the expected discounted sum of future rewards \cite{Rashid2018}.

\subsection{Successor Features}

Successor representation (SR) was introduced in \cite{Dayan1993} as an approach for separating environment dynamics from rewards in MDPs, which can be used, for example, to enable fast policy adaptation \cite{Dayan1993,Gershman2012,Momennejad2017} and task decomposition \cite{Machado2018}.
The SR defines the expected discounted future occupancy of state $s'$, given starting state $s$, action $a$, and policy $\pi$, as,
\begin{equation}\label{eqn:SR_psi}
    M^\pi(s,a,s') = \EX_\pi \left[\sum_{l=0}^\infty\gamma^{l}\mathbbm{1} [s_{t+l+1} = s']|s_t=s,a_t=a\right],
\end{equation}
where $\mathbbm{1[\cdot]}$ is one when the input argument is true. The SR effectively measures the interconnectedness of different states in the environment, i.e., the dynamics of the environment.
Given the SR, the action value function can be approximated as,
\begin{equation}
    Q^\pi(s,a) = \sum_{s'}M(s,a,s'){R}(s').
    \label{eqn:SR_Value}
\end{equation}
SR thus distinctly separates the environment dynamics, captured by $M$, from the environment reward structure, captured by $R$.

Successor features (SFs) generalizes SR to allow for the use of function approximation and applications in continuous state spaces \cite{Barreto2017,Zhang2017,Madjiheurem2019}.
The basis of SFs is that the expected one-step reward can be decomposed into a set of features, $\boldsymbol\phi(s)$, and a linear reward weighting, $\textbf{w}$, as follows,
\begin{equation}\label{eqn:SF_r}
    {R}(s) = \boldsymbol{\phi}(s)^T\cdot\textbf{w}.
\end{equation}
SFs then define the expected discounted future occupancy of features $\boldsymbol{\phi}$, given starting state $s$ and policy $\pi$, as,
\begin{equation}\label{eqn:SF_psi}
    \boldsymbol\psi^\pi(s) = \EX_\pi\left[\sum_{l=0}^\infty\gamma^{l}\boldsymbol\phi(s_{t+l+1})|s_t=s\right],
\end{equation}
where the $i$th element of $\boldsymbol\psi$ represents the expected discounting of the $i$th feature of $\boldsymbol\phi$.
The state value function can then be calculated as the following,
\begin{align}
    V^\pi(s) &= \boldsymbol\psi^\pi(s)^T\cdot\textbf{w}.\label{eqn:SF_v}
\end{align}

The original SR and SF formulations assume a fully observable MDP. 
We use SFs in our dec-POMDP setup by defining $\boldsymbol\phi(s)$ and $\boldsymbol\psi^\pi(s)$ of individual agents as $\boldsymbol\phi(\tau^i_{t})$ and $\boldsymbol\psi^{\pi^i}(\tau^i_{t})$, respectively.
For the remainder of this paper, we omit the $\pi$ notation in all global value functions and the $\pi^i$ notation in all individual value functions and SF notations for simplicity.


\section{Related Work}
The previous works most related to ours focus on decentralized coordination through careful construction of private utilities (often referred to as individual value functions).
The value decomposition network (VDN) \cite{Sunehag2018} considers the global value to be a linear summation of individual value functions.
QMIX \cite{Rashid2018} extends VDN by modeling the global value as a non-linear function of individual value functions through the use of a mixing network, with a centralized update that guarantees full factoredness.
WQMIX \cite{Rashid2020a} further extends QMIX by modifying the update step to use a weighted update over joint actions, thereby reducing QMIX's underestimates of the value approximation.
LIIR \cite{Du2019} proposes an alternative approach to improving credit assignment by reshaping individual agent objective functions with a learned intrinsic reward term. 
LICA \cite{Zhou2020} presents an actor-critic architecture that uses the mixed critic from QMIX and includes a modified policy-gradient exploration step to improve exploration near locally-optimal solutions.

These methods show promising results, but do not directly address the learnability of decentralized agents.
VDN theoretically provides infinite learnability, but only under the assumption that the global value is truly a linear sum of individual value functions, which has been empirically shown to be a weak assumption in complex games.
We aim to achieve high learnability, while maintaining high factoredness, in an effort to better handle noisy updates in decentralized settings for improved training speed, stability, and performance.
COMA \cite{Foerster2018} aims to achieve high factoredness and learnability by using an EDU critic as an advantage for a policy gradient.
QTRAN \cite{Son2019} also aims for high learnability by using a similar counterfactual joint action value function to learn agent-specific contributions to a task.
However, COMA and QTRAN require a function approximation of the joint state action value.
We address learnability in a manner that removes this requirement for better scaling in MAS with complex interactions among many agents.

Other works achieve coordination through reward redistribution amongst individual agents \cite{devlin2011theoretical,devlin2014potential,tang2021discovering}, but require domain-specific knowledge to define the reward shaping mechanisms.
SFs have also been applied to MARL in previous works.
\cite{Gupta2019a} uses SFs in lieu of traditional value networks, allowing for efficient policy transfer to new reward structures.
However, \cite{Gupta2019a} assumes individual rewards are given to agents; we instead focus on environments where a global reward is given with no prior definition of individual rewards.
\cite{gupta2021uneven} uses SFs within a VDN structure and uses its decomposition of dynamics and rewards for better value estimation under dynamic policies when learning. Our method instead proposes an entirely new line of work in which we leverage the linear feature composition of SFs to disentangle an individual agent's value from all other agents in the global value function.

\section{Our Approach}
\begin{figure*}[t]
   \centering
   \includegraphics[width=1.0\textwidth]{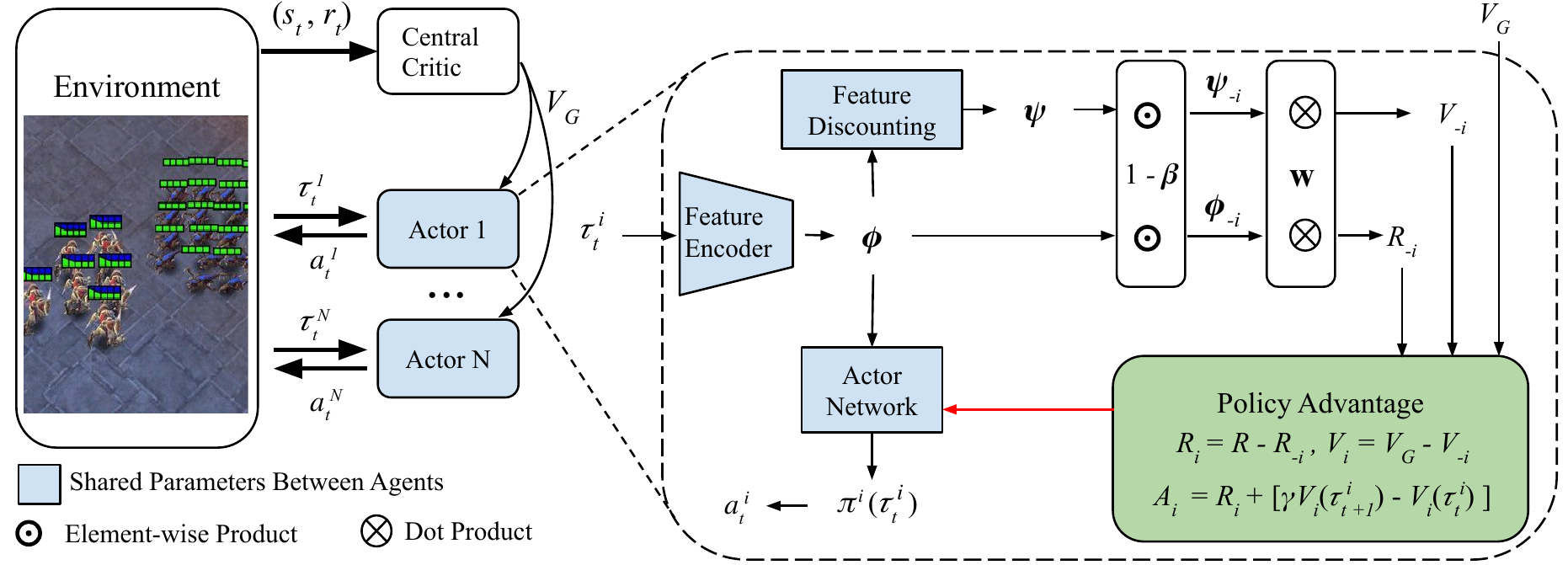} 
   \caption{The architecture used to implement our DISSC framework. Our framework uses SFs to learn a disentanglement vector $\beta$ that allows us to estimate individual agent value and reward functions, $V_i$ and $R_i$, respectively. We use $V_i$ and $R_i$ to train decentralized agents, supported by a central critic used to estimate the global value function $V_G$. We optimize $\beta$ for learnability, which we estimate using SFs, to improve training stability in complex environments with many agents.}
   \label{fig:method}
   \vspace{-10pt}
\end{figure*}
We present the \underline{Dis}entangled \underline{S}Fs for \underline{C}oordination (DISSC) framework, which leverages SFs to disentangle the impact of an individual agent from the global value function. We use this disentanglement to improve coordination in MARL by compactly representing the EDU private utility and learnability.
Our framework is composed of three key ideas: (1) an SF-EDU formulation that uses SFs to represent the marginalization term in the EDU for unstructured tasks; (2) a process for learning an SF disentanglement vector that is optimized for learnability and used to calculate our SF-EDU; and (3) a CTDE architecture that efficiently implements these techniques. Our overall architecture is summarized in \Cref{fig:method}.

\subsection{SF-EDU Private Utility Formulation}
\label{ss:modified_edu}

A challenge with using the EDU private utility is efficiently calculating the marginalization term, i.e., $\EX_{z_i}[G(\textbf{z})|\textbf{z}_{-i}]$.
Rather than marginalizing $G(\textbf{z})$ with a joint move, we instead present a modified EDU formulation that uses SFs to marginalize $G(\textbf{z})$ with $\boldsymbol\psi$.
We define our SF-EDU private utility as,
\begin{align}
    EDU_i &\approx V_G(s) - \EX_{\pi^i}[Q_G(\tau^i_t,a^i)|\boldsymbol\psi_{-i}] \label{eqn:modified_edu1}\\
    &\approx V_G(s) - V_{-i}(\tau^i_t) \label{eqn:modified_edu2}\\
    &\approx V_G(s) - \boldsymbol\psi_{-i}^T\cdot \textbf{w} \,, \label{eqn:modified_edu}
\end{align}
where $\boldsymbol\psi_{-i}$ are the SFs associated with the contribution of all agents other than agent $i$ to the global value.
We define the second term in \Cref{eqn:modified_edu1} with respect to local agent information due to our dec-POMDP formulation.
Similar to the original EDU formulation in \Cref{eqn:EDU_Tumer}, the second term in \Cref{eqn:modified_edu} represents the expected global value (or utility) associated with all agents other than $i$.
We implement our formulation by calculating the global value $V_G$ with a central critic and the marginalizing term using a decentralized critic.
We then define $V_i = EDU_i$ to calculate a generalized advantage estimator $A_i=R_i(\tau^i_{t}) + [\gamma V_i(\tau^i_{t+1})-V_i(\tau^i_t)]$ for training our decentralized actor policies.
We discuss our implementation in more detail in \Cref{ss:architectures}.

The intuition behind SF-EDU is that certain features in an agent’s observation are likely not relevant to that agent’s contribution to the global value, and instead mostly contain noise associated with other agent contributions to the global value (see \Cref{fig:explainability} for a visual example of this idea). We represent those non-relevant features as $\boldsymbol\psi_{-i}$. In principle, one could identify these non-relevant features using any latent features. However, our choice to use SFs allows us to directly relate the non-relevant features to a marginalized value function (i.e., $V_{-i}$), due to the unique separation property of SFs, where the value function can be decomposed into the product of a feature representation and a linear weighting, as shown in \Cref{eqn:SF_v}.

\subsection{SF Disentanglement Vector for Marginalizing Global Reward and Value}
\label{ss:sf_filter}

We estimate $\boldsymbol\psi_{-i}$ by learning an SF disentanglement vector, $\boldsymbol\beta$, that filters an agent's SF encoding to only focus on features relevant to that agent's contribution to the global value and reward.
More formally, we implement $\boldsymbol\beta$ as an element-wise rescaling vector that operates on an agent's features $\boldsymbol\phi$ and SFs $\boldsymbol\psi$ as follows,
\begin{align}
    \boldsymbol\phi_{-i} = \boldsymbol\phi\odot(1 - \boldsymbol\beta),\label{eqn:rescaling_phi}\\
    \boldsymbol\psi_{-i} = \boldsymbol\psi\odot(1 - \boldsymbol\beta),\label{eqn:rescaling_psi}
\end{align}
where $\odot$ denotes an element-wise product. We use the rescaled SFs $\boldsymbol\psi_{-i}$ to estimate $V_{-i}$ for calculating our SF-EDU, as shown in \cref{eqn:modified_edu}. Similarly, we leverage \Cref{eqn:SF_r} to use the rescaled features $\boldsymbol\phi_{-i}$ to estimate the individual reward $R_i$ associated with agent $i$ (i.e., the reward generated solely by agent $i$) as follows,
\begin{equation}
    R_i = R - R_{-i} = R - ({\boldsymbol\phi_{-i}}^T \cdot \textbf{w}).
    \label{eqn:ind_r}
\end{equation}
This quantity is used in the generalized advantage estimator for training individual agent policies.
We focus on $\boldsymbol\phi_{-i}$ and $\boldsymbol\psi_{-i}$ rather than their complements because we find them to improve stability during training, likely due to better initialization.

We learn $\boldsymbol\beta$ by optimizing it with respect to the learnability of SF-EDU.
The intuition behind this idea is based on the fact that SF-EDU aims to approximate EDU, which achieves high learnability through its marginalization term. Since we estimate this marginalization term using $\boldsymbol\psi_{-i}$, which is calculated from $\boldsymbol\beta$, by optimizing $\boldsymbol\beta$ such that SF-EDU has high learnability, we ensure that $\boldsymbol\beta$ effectively filters an agent's observation to focus on features relevant to its contribution to the global value function.

However, calculating learnability is challenging in complex environments.
We therefore use SFs to derive a novel representation of learnability that can be estimated in such environments.
Beginning with the original learnability equation and re-formulating it to use value functions as utilities, we represent the learnability $\lambda_{i}(\textbf{a})$ of an agent's individual value function as follows,
\begin{align}
    \lambda_{i}(\textbf{a}) &= \frac
        {\EX_{a_t^{i'}} \left[ |Q_i(\tau^i_t, \textbf{a}_t)-Q_i(\tau^i_t,\textbf{a}_t^{-i} + a_t^{i'})| \right]}
        {\EX_{\textbf{a}_t^{-{i'}}} \left[ |Q_i(\tau^i_t, \textbf{a}_t)-Q_i(\tau^i_t,\textbf{a}_t^{-i'}+ a_t^i)| \right]} \label{eqn:learnability_SF1}\\
    &\approx \frac
        {\EX_{\pi^i,\tau^i_{t\rightarrow t+1}} \left[ |
            \boldsymbol\psi(\tau^i_t)\textbf{w}
            -\boldsymbol\psi(\tau^i_t)\textbf{w}^-
            -\boldsymbol\psi(\tau^i_{t+1}) \textbf{w}^+
        |\right]}
        {\EX_{\pi^{-i},\tau^i_{t\rightarrow t+1}} \left[ |
            \boldsymbol\psi(\tau^i_t)\textbf{w}
            -\boldsymbol\psi(\tau^i_{t+1}) \textbf{w}^-
            -\boldsymbol\psi(\tau^i_t) \textbf{w}^+ |
        \right]} \\
    &\approx \frac
        {\EX_{\pi^i,\tau^i_{t\rightarrow t+1}}[|\left(\boldsymbol\psi(\tau^i_t)-\boldsymbol\psi(\tau^i_{t+1})\right)\cdot \textbf{w}^+|]}
        {\EX_{\pi, \tau^i_{t\rightarrow{t+1}}}    [|\left(\boldsymbol\psi(\tau^i_t)-\boldsymbol\psi(\tau^i_{t+1})\right)\cdot \textbf{w}^-|]}, \label{eqn:learnability_SF}
\end{align}
where,
\begin{align}\label{eqn:w_definition} 
\begin{split}
    w^+ &= \boldsymbol\beta\odot \textbf{w}\\
    w^- &= (1-\boldsymbol\beta)\odot \textbf{w}.
    \end{split}
\end{align}
Similar to the original learnability, our representation measures the ratio of the expected value (or utility) change over variations of agent $i$'s actions relative to the expected value change over all other agent actions.
We approximate the value over variations of agent $i$'s actions $\left[ Q_i(\tau^i_t,\textbf{a}_t^{-i} + a_t^{i'}) \right]_{a_t^{i'}}$ as the transition of value between two consecutive states $\left[ \boldsymbol\psi(\tau^i_t)\cdot \textbf{w}^- + \boldsymbol\psi(\tau^i_{t+1})\cdot \textbf{w}^+ \right]_{\pi^i, \tau_{t\rightarrow t+1}^i}$; we use a similar approximation for the value over variations of all other agent actions.
These approximations are based on the assumption of independent features in $\boldsymbol\psi$ and the fact that we define $\boldsymbol\psi_i$ and $\boldsymbol\psi_{-i}$ to be complementary vectors, as shown in \Cref{eqn:w_definition}. We enforce complementarity by constraining $\beta \in [0,1]$ in the update process, further described in \Cref{alg:alg}. 
We also calculate the expectation in the denominator of \Cref{eqn:learnability_SF} over the complete joint action rather than the joint action of agents other than $i$, resulting in \Cref{eqn:learnability_SF} being a conservative estimate of learnability, since our computation includes the influence of agent $i$'s action.

Based on our derived representation for learnability, shown in \Cref{eqn:learnability_SF}, we update $\boldsymbol\beta$ in a manner that maximizes learnability using the following loss,
\begin{align}\label{eqn:loss_learnability}
    L_{\lambda, \boldsymbol\beta} = \sum_t\left[\sum_{a_t^i} \pi_i(a_t^i|\tau^i_t)
        \left\|\left(\boldsymbol\psi(\tau^i_t)-\boldsymbol\psi(\tau^i_{t+1})\right)\cdot \textbf{w}^+\right\|\right] \nonumber\\
        -c_\lambda\left\|\left(\boldsymbol\psi(\tau^i_t)-\boldsymbol\psi(\tau^i_{t+1})\right)\cdot \textbf{w}^-\right\|.
\end{align}
The numerator and denominator of \Cref{eqn:learnability_SF} are optimized separately for numerical reasons and $c_\lambda$ is introduced to mitigate the impact of magnitude differences between those quantities during training.

\subsection{CTDE Architecture}
\label{ss:architectures}

We implement our framework using a CTDE architecture, building from recent work suggesting that similar architectures encourage coordination given a global reward \cite{Corder2019,Lowe2017}.
While we include centralized training in this work, our framework can be applied to fully decentralized architectures with little modification.

Our central critic builds from the formulation used by COMA \cite{Foerster2018}, in that it estimates the global value and sends this value to individual agents for their EDU calculation.
However, we estimate the global state value $V_G$ instead of the global action value $Q_G$ to better handle large-scale MAS.
Our decentralized controllers implement their own internal actor-critic architectures.
We train the decentralized critics using SFs, where we use the learned $\boldsymbol{\phi}$ and $\boldsymbol{\psi}$ to estimate the global reward $R$ and state value $V_G$ with \Cref{eqn:SF_r,eqn:SF_v}, respectively.
We optimize individual agents' policies, $\pi^i(a^i|\tau^i_t)$, with the proximal policy optimization (PPO) algorithm  \cite{Schulman2017a}, using our SF-EDU to calculate the advantage in lieu of the traditional advantage function (as discussed in \Cref{ss:modified_edu}).

Our architecture uses a shared SF encoder for all agents.
This shared encoder is motivated by existing work demonstrating benefits of weight sharing \cite{Dutta2005,Foerster2016,OliveiraSouza2019} and based on a hypothesis that using SFs, we can model spatial and dynamic features in an environment in a manner that is common to all agents regardless of their type.
The weight-sharing significantly reduces network complexity and paves the way towards efficiently utilizing experiences in heterogeneous MARL training.

\subsection{Successor Features Implementation} \label{sec:SF}

We train the components of our SF critic, the feature encoder, and feature discounting using standard SF losses \cite{Kulkarni2016,Zhang2017}.
We first train the feature encoder to recover a set of features, $\boldsymbol\phi$, that encodes information from local surroundings using the following reward estimation loss,
\begin{equation}\label{eqn:loss_reward}
    L_{\text{reward}, \boldsymbol\theta_{\boldsymbol\phi}} = \left\| r_t - \boldsymbol\phi^{\theta_{\boldsymbol\phi}}(\tau^i_t)^T \cdot \textbf{w} \right\|^2.
\end{equation}
Using this loss alone, however, often does not generate a dense enough feature set for stable training, particularly in sparse reward environments.
Motivated by \cite{Barreto2018}, we address this issue by implementing auxiliary training networks to predict the next state and increase the amount of encoded information seen by $\boldsymbol\phi$, using the following prediction loss,
\begin{equation}\label{eqn:loss_prediction}
    L_{\text{pred}, \boldsymbol\theta_{\boldsymbol\phi}} = \left\|{o^i_{t+1}} - \mathcal{D}(\boldsymbol\phi^{\theta_{\boldsymbol\phi}}(o^i_t),a^i_t)\right\|^2,
\end{equation}
where $\mathcal{D}(\boldsymbol\phi({s^i_t}),a^i_t)$ is a decoder network that predicts the next state observation. The feature discounting network is trained to discount the learned features with the following SF loss, 
\begin{equation}\label{eqn:loss_TD_psi}
    L_{\text{SF},\theta_{\boldsymbol\psi}} = \left\| \boldsymbol\phi(\tau^i_{t+1}) + \gamma \boldsymbol\psi^{\theta_{\boldsymbol\psi}}(\tau^i_{t+1}) - \boldsymbol\psi(\tau^i_t)\right\|^2,
\end{equation}
which is the TD-error of the SFs $\boldsymbol\psi$ generalized into a Bellman equation. 
Our overall algorithm is summarized in \Cref{s:algorithm}.


\section{Results}
\label{sec:results}
We test our approach through a set of experiments using the open-source Multi-Agent Particle (MAP) environments \cite{Lowe2017}, SMAC \cite{Samvelyan2019}, and a capture-the-flag (CtF) environment \cite{Stralen2020}.
We compare to the following published MARL baselines: COMA \cite{Foerster2018}, QMIX \cite{Rashid2018}, QTRAN \cite{Son2019}, WQMIX \cite{Rashid2020}, LIIR \cite{Du2019}, and LICA \cite{Zhou2020}. We tuned baselines for environments (and individual maps in SMAC) that they were not originally demonstrated on by running several experiments over common hyperparameters, primarily learning rate and batch size, and method specific variables, such as $\alpha$ in WQMIX. 
We also extended the $\epsilon$-greedy exploration to 1M episodes for $Q$-learning baselines to improve exploration in hard SMAC maps and CtF, as suggested in \cite{Rashid2020}.
Note that our method and all baselines implemented use a CTDE architecture.
All figures plot the mean performance with one standard deviation shaded.

Similar to existing MARL architectures \cite{Rashid2018,Rashid2020,Du2019,Zhou2020}, DISSC used an MLP feature encoder with an LSTM in the Predator Prey and SMAC environments.
We used MLPs for all other networks.
For the CTF environment, which uses an image-like input, convolutional layers were added to the head of the feature encoder to capture spatial relationships.
All networks were optimized with an Adam optimizer with a learning rate of 1E-4 based on the losses described in \Cref{eqn:loss_reward,eqn:loss_prediction,eqn:loss_TD_psi}.
We set $c_\lambda = 0.5$ for the results shown; our testing found our results to have little sensitivity to the value of $c_\lambda$.
Code is available at \textbf{\url{https://github.com/Tran-Research-Group/DISSC}}.

\subsection{Multi-Agent Particle Environments} \label{ss:map_exp}
\begin{wrapfigure}{R}{0.55\textwidth}
\vspace{-0.1in}
  \includegraphics[width=0.55\textwidth]{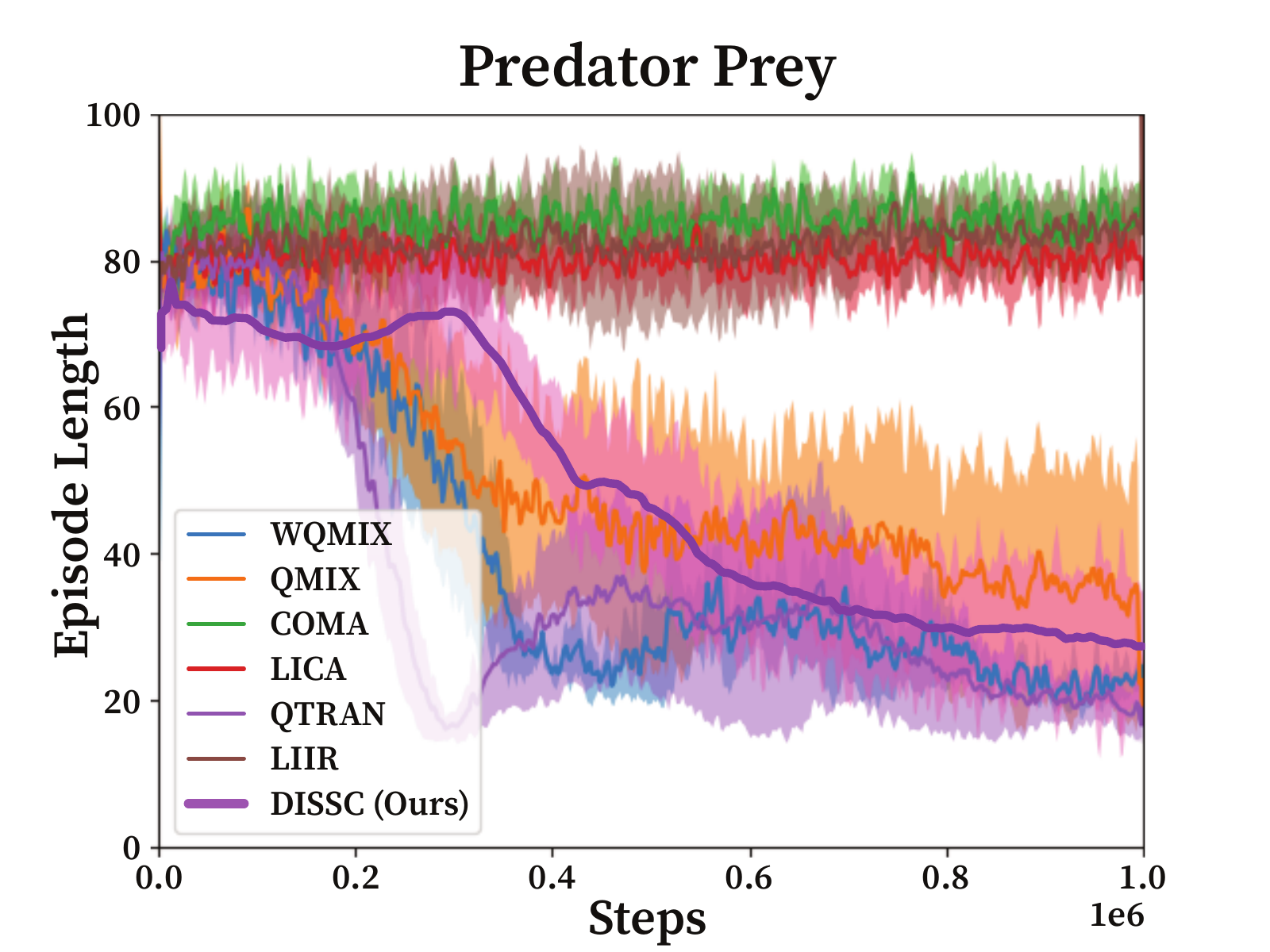} 
  \caption{Training convergence plots in the predator-prey environment, over 10 replicates. Most methods solve the task with some variation in learning speed. Lower episode length indicates better performance.}
  \label{fig:PredPrey}
  \vspace{-20pt}
\end{wrapfigure}
We first consider the predator-prey environment from the MAP test suite \cite{Lowe2017}. 
The goal of predator-prey is for three predators to capture a randomly moving agent as fast as possible. 
The environment requires coordination to efficiently and quickly solve the task.


\Cref{fig:PredPrey} shows results for this environment.
We observe that several methods, including DISSC, are able to sufficiently solve the task by minimizing the time taken to capture the prey.
COMA, LICA, and LIIR struggled with this environment.
While DISSC does converge to optimal performance, we note that it is slower than some alternatives.
We believe that this is due to our added complexity of learning SFs to model the value function; for this simple environment, that complexity does not outweigh the benefit of our disentanglement approach.

\begin{figure*}[t]
  \centering
     \includegraphics[width=1.0\textwidth]{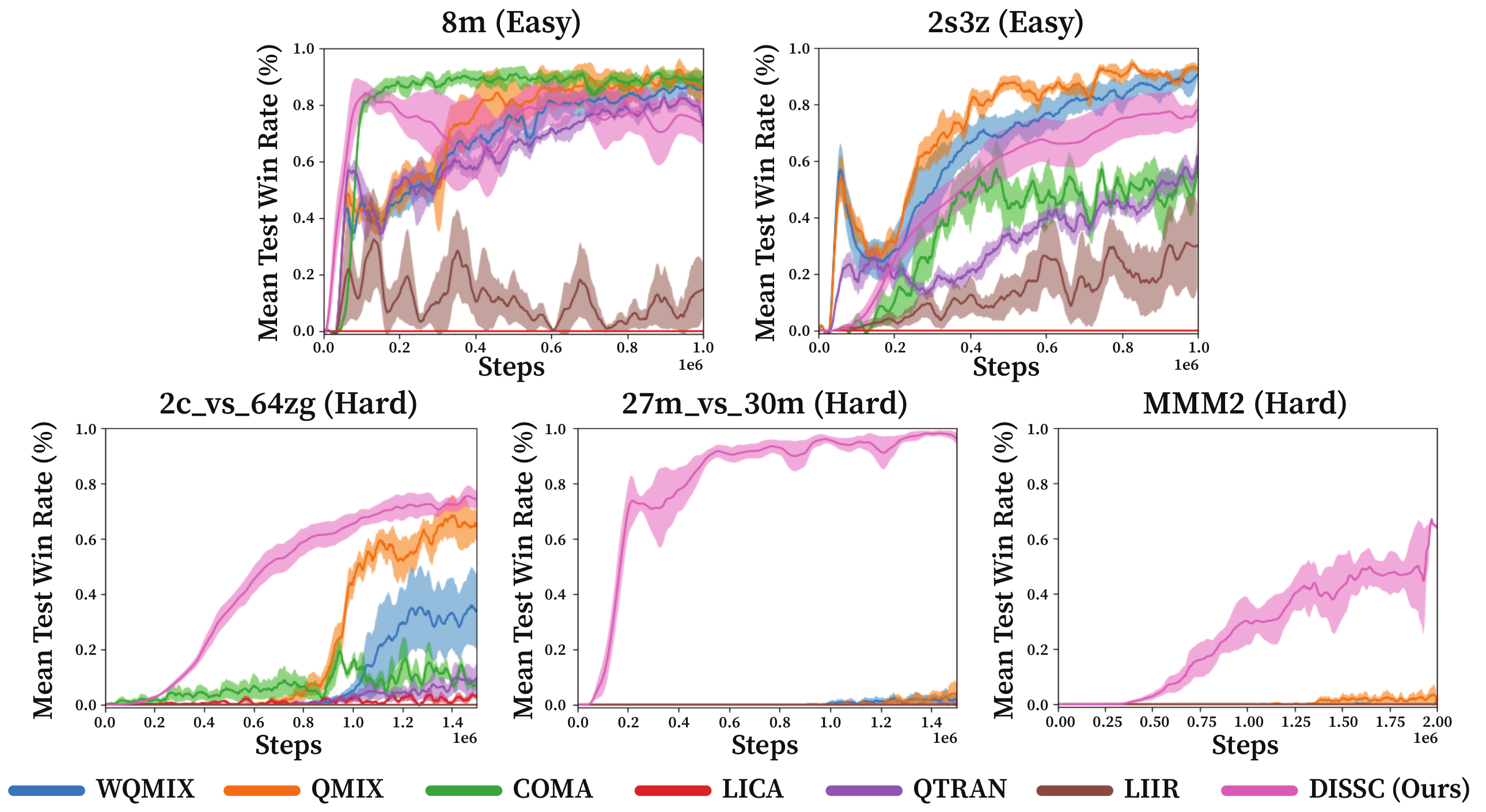}
  \caption{Training convergence plots in five SMAC maps, calculated over five replicates. A higher win-rate indicates better performance. In easier games, 8m and 2s3z, we match performance of the baselines. In harder games (which control more agents), our method has significantly faster learning times and higher performance.}
  \label{fig:SC2_Resuls}
  \vspace{-20pt}
\end{figure*}

\subsection{Starcraft Multi-Agent Challenge} \label{ss:sc2_exp}

We also consider the SMAC environment, a Starcraft II mini-game that focuses on micromanaging a set of agents in combat.
We consider several maps ranging in difficulty and heterogeneity of the controlled agents.
Easy maps have smaller numbers of controlled units and fight an equal number of enemies (8 marines vs. 8 marines in 8m, 2 stalkers and 3 zealots vs 2 stalkers and 3 zealots in 2s3z).
Hard maps either place the agents at a numbers disadvantage (2 colosi vs. 64 zerglings in 2c\_vs\_64zg), control a large number of units (27 marines in 27m\_vs\_30m), or control highly heterogenous units (1 Medivac, 2 Mauraders, \& 7 Marines in MMM2).

\Cref{fig:SC2_Resuls} shows our results for various SMAC maps.
For easy maps, we see that nearly all methods are able to attain high levels of performance, including DISSC.
Our method shows fast learning in 8m but slower initial learning in 2s3z, again suggesting that the benefits of learning SFs towards learning speed may be reduced for simple environments.

In harder maps, we begin to observe the performance benefit of our method.
For the 2c\_vs\_64zg map, only DISSC, QMIX, and WQMIX are able to obtain a non-trivial positive win-rate; all other methods essentially fail to solve the task.
However, amongst these three, our method shows significantly faster learning, the highest converged performance, and the lowest variance in win rate.
As we move to maps where more agents are controlled (i.e., 2c\_vs\_64zg, MMM2, 27m\_vs\_30m), we see that our method is the only one able to solve the task.
We expect that these performance gains are due to the learning stability provided by our method through its direct optimization of learnability, which is most beneficial in complex environments with many agents.
There may also be some benefit incurred by our architecture, since we use a simplified central critic, $V_G=f(s)$, instead of, for example, complex mixing networks used by QMIX and WQMIX which scale linearly with the number of agents, $Q_G=f(s,V_1,..., V_\mathcal{I})$.
Our architecture thus better scales to large MAS like 27m\_vs\_30m and MMM2.

\begin{figure*}[t]
  \centering
     \includegraphics[width=1.0\textwidth]{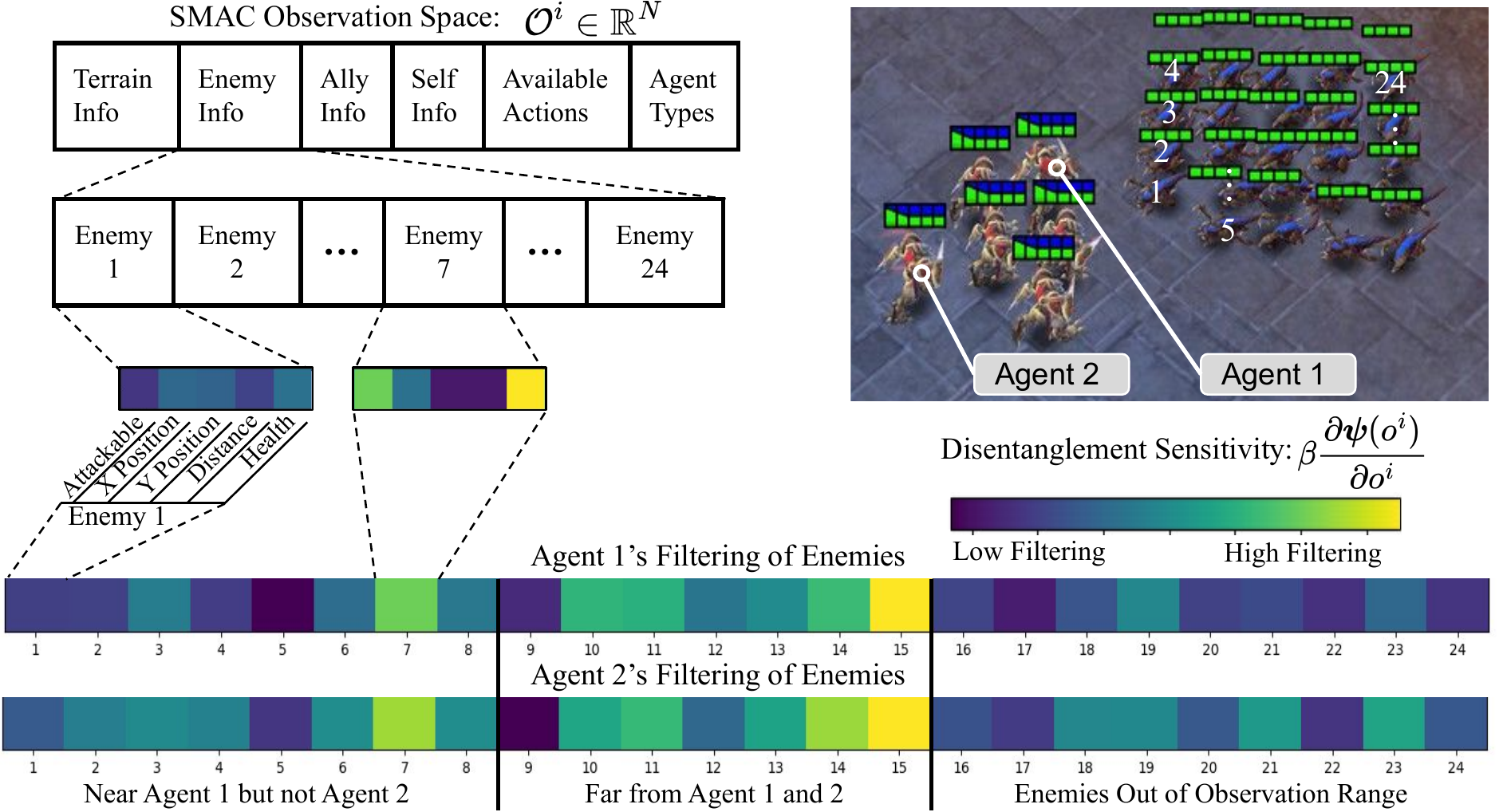}
  \caption{Effect of the disentanglement vector $\boldsymbol\beta$ on agent learning. Agent observations in SMAC are a vector with information about current position, enemies, allies, and available actions. We use Grad-CAM \cite{Selvaraju2020} to correlate the sensitivity of the feature space, $\boldsymbol\psi$, with respect to the observation, $o^i$. The gradient, ${\partial \boldsymbol\psi}/{\partial o^i}$, can then be multiplied by the disentanglement vector, $\boldsymbol\beta$, to understand which observation variables are the most heavily filtered during disentanglement. The breakout shows Agent 1 filters Enemy 1 less than Enemy 7, primarily by their ``Attackable'' and ``Health'' variables. This filtering occurs because Agent 1 is closest to Enemy 1 and can attack the enemy to generate reward by damaging them. The bottom breakout shows the summation of the filtering across all enemies for Agents 1 and 2 and supports the trend that enemies further away from agents are more heavily filtered.}
  \label{fig:explainability}
  \vspace{-14pt}
\end{figure*}

To better understand our method, we also visualize the effect of our disentanglement vector $\boldsymbol\beta$ on individual agents, shown in \Cref{fig:explainability}.
We accomplish this by using a Grad-CAM-like \cite{Selvaraju2020} approach to correlate the effect of observations to the latent space $\boldsymbol\psi$.
More specifically, we calculate a correlation, ${\partial \boldsymbol\psi}/{\partial o^i}$, that captures the effect of changing the observation on the feature space.
We then multiply this correlation by  $\boldsymbol\beta$ to evaluate which elements of the observation space are being filtered by $\boldsymbol\beta$.

We observe two interesting trends from this analysis.
First, we see that $\boldsymbol\beta$ heavily filters specific elements of the observation space, primarily the ``Enemy Health'' and ``Enemy is Attackable'' variables for nearly every enemy.
This result is encouraging, since rewards in SMAC are given for damaging enemy units, suggesting that $\boldsymbol\beta$ learns to assign value based on whether or not an enemy is within attack range and its health.
Second, we see that agents filter different features depending on their position with respect to enemies, as agents far away from enemies filter enemy variables more heavily than agents closer to enemies.

\subsection{Capture the Flag (CtF)}
We further test our approach in a CtF environment to understand its performance in a sparse reward setting. Here, two teams of agents compete against one another to capture their opponent's flag while simultaneously defending their own flag.
Agent interactions focus on flag capture and stochastic engagements in which agents can be temporarily removed from gameplay; these engagements depend on various factors such as agent strength, proximity to other agents, and home field advantage.
Agents are randomly re-spawned within their own territory after being removed from gameplay.
Flags are randomly re-spawned within their team's territory immediately after being captured.
We use an open-source implementation of the game \cite{Stralen2020} that operates in a 2D grid-world.

\begin{figure*}[t]
  \centering
    \includegraphics[width=0.90\textwidth]{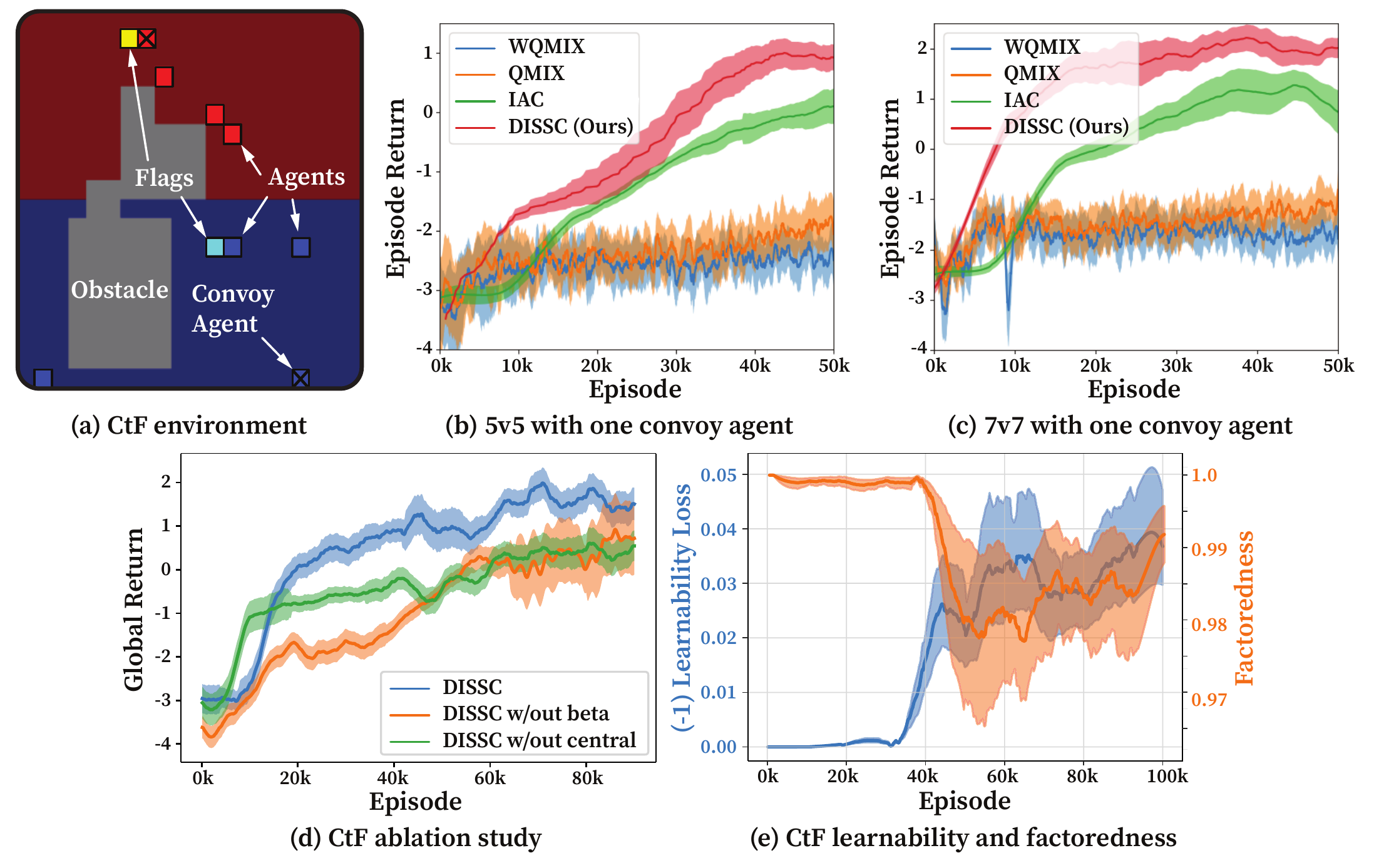}
  \caption{(a) The CtF environment. (b)-(c) Training convergence plots for our considered methods in two CtF settings, calculated over five replicates. Our method shows improved training time, converged performance, and stability over baselines. We also show (d) ablation results and  (e) learnability and factoredness improvements during training.}
  \vspace{-15pt}
  \label{fig:ctf}
\end{figure*}

We compare to WQMIX and QMIX (since they show the most promise among baselines in SMAC hard environments) in this experiment, along with a vanilla independent actor-critic (IAC) baseline.
We experiment in 5-vs-5 and 7-vs-7 games, where teams are composed of two types of agents: slow, strong ``convoy'' agents and fast, weak ``normal'' agents.
\Cref{fig:ctf} shows that our method, DISSC, significantly outperforms QMIX and WQMIX, with IAC falling somewhere in the middle.
We expect that the poor performance of QMIX and WQMIX is due to convergence to a defensive policy where the agents remain on their own territory to limit their chances of being killed.
This sub-optimal policy likely occurs due to limited exploration in this sparse reward environment, a known issue with many $Q$-learning methods.

\Cref{fig:ctf}d also contains results from an ablation study of DISSC, comparing the effects of our $\boldsymbol\beta$ disentanglement vector and our centralized critic on model performance.
We see that removing $\boldsymbol\beta$ reduces training stability, supporting our assertion that our modified EDU improves training stability through its high learnability.
We also see that removing the central critic, and instead allowing individual agents to evaluate the global value, results in the lowest converged global return within our model variants, suggesting that local SF critics struggle to estimate the global value. 
Local critics likely struggle because agents only have access to their own individual observations during training. 

We also investigate how factoredness and learnability change during training to better understand our SF-EDU learning process, as shown in \Cref{fig:ctf}e.
We calculated factoredness using \Cref{eqn:factoredness_Tumer} and learnability using \Cref{eqn:learnability_SF}.
We see that during training, DISSC is able to explicitly increase learnability of individual agents, though there appears to be a slight tradeoff between factoredness and learnability.
This tradeoff is likely due to our approximation of the EDU, but is minimal relative to changes in learnability.
These results suggest that learnability and factoredness may naturally conflict in some environments and therefore require simultaneous optimization of both in such settings.

\section{Conclusions and Future Work}
\label{sec:conclusions}

A critical challenge in MARL is ensuring coordination among decentralized agents in unstructured and complex tasks. We show that SFs can be used to disentangle the impact of individual agents on the global value function, from the impact of all other agents. This disentanglement allows us to compactly represent a individual value function that has high factoredness and high learnability such that we can train decentralized agents in a coordinated and stable manner. We implement our approach in a CTDE architecture and demonstrate that it shows improved training time and performance over alternatives in several multi-agent environments.

For future work, we suggest exploring learning SF disentanglements that account for specific agent types. While our results suggest such heterogeneity is not required, further developments may be able to leverage agent-specific disentanglements to identify potential roles or specializations for a given task.
This notion is also inspired by recent biological discoveries in mammals that have shown cell specialization to encode environment information in a semi-structured format, instead of an unknown latent space \cite{Tsao2013, Schultz2015,Stachenfeld2017,Montchal2019}.

\begin{ack}
This work was supported by ONR Grant N00014-20-1-2249 and ARL Contract W911NF2020184.
\end{ack}

\clearpage 

\bibliographystyle{abbrv}
\bibliography{Bib_Seung,NVS_Bib,HT-library,misc}
\clearpage
\clearpage
\appendix

\section{Algorithm} \label{s:algorithm}

    \begin{algorithm}[h]
    \SetAlgoLined
    Let $i\in \mathcal{I}$ : index for agent, $d\in\textbf{N}$ : index for type of agent\;
    Initialize $\Theta$, $\theta^d$, $\pi^d$, $\psi^d$, $\boldsymbol\beta^d \leftarrow 1$\;
    Initialize trajectory buffer $D_{\text{central}}$, $D^d_{\text{decentral}}$ \;
    \Repeat{
        Reset environment\;
        Bootstrap action $\pi(s^1_0)$, $\pi(s^2_0)$, $...$\;
        \While{$t< \text{max step}$}{
            Obtain observations $s_t$, $s^1_t$, $s^2_t$, $...$ from environment\;
            Obtain actions $\pi(s^1_t)$, $\pi(s^2_t)$, $\pi(s^3_t)$, $...$\;
            Execute actions ($a_1$,$a_2$,$a_3$,$a_4$,$...$)\;
            Store $D_\text{central}\leftarrow(s_t, r_t)$\;
            Store $D^d_\text{decentral}\leftarrow(s^i_t, r_t, a^i_t)^d$\;
            \If{$|D_\text{central}|<\text{central batch size}$}{
                Sample ($r_t$, $V(s_t)$) $\forall s_t\in D_\text{central}$\;
                Compute TD-Target: $TD_\text{central}=r_{t+1}+\gamma V^{-\Theta}(s_{t+1})$\;
                Update Central Critic: $\Theta\leftarrow\Theta-c_\text{central}\nabla_\Theta(TD_\text{central}-V(s_t))^2$\;
                Reset trajectory buffer $D_{\text{central}}$\;
            }
            \If{$|D_\text{decentral}|<\text{decentral batch size}$}{
                Sample $\pi^d$, $r_t$, $\phi_t$, $\psi_t$, $V_t$, $\forall s_t\in D_\text{decentral}$\;
                Compute TD-Target: $TD=r_{t+1}+\gamma V^{-\theta}(s_{t+1})$\;
                Compute TD-SF: $TD_\psi=\phi_{t+1}+\gamma \psi^{-\theta_\psi}(s_{t+1})$\;
                Update Agent policy: $\theta_\pi\leftarrow\theta_\pi-c_\pi\nabla_{\theta_\pi}L_{ppo}$\;
                Train SF Representation:\;
                $\theta_\psi\leftarrow\theta_\psi-c_\psi\nabla_{\theta_\psi}(TD_\psi-\psi(s_t))^2$\;
                $\theta_\phi\leftarrow\theta_\phi-c_\phi\nabla_{\theta_\phi}(R_t-\phi\cdot w)^2$\;
                Update Learnability rescaling of agents: $\beta\leftarrow\beta-c_\lambda\nabla_\beta L_{\lambda,\beta}$\;
                $\beta = max(0.0, \beta)$\;
                Reset trajectory buffer $D^d_{\text{decentral}}$\;
            }
            \If{environment done}{
                Break
            }
        }
    }
    \caption{DISSC Algorithm}\label{alg:alg}
    
    \end{algorithm}

\section{Experiments}\label{apd:Experiments}

All SMAC experiments were performed using SC2.4.10. Performance is not consistent between versions, and several previous works \cite{Rashid2020} use SC2.4.6.


\end{document}